\documentclass[sigconf]{acmart}

\AtBeginDocument{%
  }

\setcopyright{acmlicensed}
\copyrightyear{2024}
\acmYear{2024}
\setcopyright{rightsretained}
\acmDOI{10.1145/3677052.3698611}

\acmConference[ICAIF '24]{5th ACM International Conference on AI in Finance}{November 14--17, 2024}{Brooklyn, NY, USA}
\acmBooktitle{5th ACM International Conference on AI in Finance (ICAIF '24), November 14--17, 2024, Brooklyn, NY, USA}
\acmISBN{979-8-4007-1081-0/24/11}

\usepackage{indentfirst}
\usepackage{algorithm}
\usepackage{algpseudocode}
\usepackage{multirow}

\begin{document}

\title{Dynamic Pricing in Securities Lending Market: Application in Revenue Optimization for an Agent Lender Portfolio}

\author{Jing Xu}
\authornote{Both authors contributed equally to this research.}
\orcid{0009-0008-1860-2057}
\affiliation{%
  \institution{J.P. Morgan Quantitative Research}
  \city{New York}
  \state{New York}
  \country{USA}
}
\email{jing.xu@chase.com}

\author{Yung-Cheng Hsu}
\authornotemark[1]
\orcid{0009-0009-8899-9886}
\affiliation{%
  \institution{J.P. Morgan Quantitative Research}
  \city{New York}
  \state{New York}
  \country{USA}
}
\email{yung-cheng.hsu@jpmchase.com}

\author{William Biscarri}
\authornote{Author served as an advisor with part of article composition.}
\orcid{0009-0005-9768-975X}
\affiliation{%
  \institution{J.P. Morgan Quantitative Research}
  \city{New York}
  \state{New York}
  \country{USA}
}
\email{william.biscarri@jpmorgan.com}

\renewcommand{\shortauthors}{Jing Xu, Yung-Cheng Hsu, and William Biscarri}

\begin{abstract}
  Securities lending is an important part of the financial market structure, where agent lenders help long term institutional investors to lend out their securities to short sellers in exchange for a lending fee. Agent lenders within the market seek to optimize revenue by lending out securities at the highest rate possible. Typically, this rate is set by hard-coded business rules or standard supervised machine learning models. These approaches are often difficult to scale and are not adaptive to changing market conditions. Unlike a traditional stock exchange with a centralized limit order book, the securities lending market is organized similarly to an e-commerce marketplace, where agent lenders and borrowers can transact at any agreed price in a bilateral fashion. This similarity suggests that the use of typical methods for addressing dynamic pricing problems in e-commerce could be effective in the securities lending market. We show that existing contextual bandit frameworks can be successfully utilized in the securities lending market. Using offline evaluation on real historical data, we show that the contextual bandit approach can consistently outperform typical approaches by at least 15\% in terms of total revenue generated.
\end{abstract}





\maketitle

\section{Introduction}
A fundamental problem in quantitative finance is the design and implementation of automated data-driven processes to operate within financial markets. A particular area of interest is automated trading, where the goal is to create systems capable of interacting with a market without human intervention. This problem has been, and continues to be, extensively studied in multiple markets, but particular focus has been on markets with a traditional limit order book structure, such as cash equities \cite{YuriyN, ZZ}, futures \cite{GuhyukChung}, and foreign exchange \cite{AntonioRiva, JCar}.

One market that has seen comparatively less attention is the securities lending (SL) market, which is an important, yet often overlooked, component of the modern financial system. In the SL market, agent lenders which hold securities on behalf of customers compete with one another to lend out these securities to other investors, who typically seek to borrow these securities for the purposes of shorting. Much past work has focused on analyzing the SL market primarily from an auction or game theoretic perspective \cite{EmilyDiana, MahdiNez, ShuaiC}.

The usual goal of agent lenders operating within the SL market is to maximize the total amount of revenue received from lending out securities to borrowers. Since there is no difference between any given security borrowed from different agent lenders, borrowers have no preference for one agent lender over another except in the cost of borrowing being offered by each lender. Furthermore, the number of available shares of any given security available to lend in the market often far surpasses the demand of that security to be borrowed. As a result, agent lenders must carefully set the cost of borrowing the securities that they hold to optimize the total amount of revenue they receive.

Traditionally, agent lenders have used rule-based logic derived from business intuition in combination with human judgement from manual traders to set the rate for borrowing a security. While this approach can be effective, the hard-coded rules are often static and may reflect observations or patterns that were valid in the past, but which no longer hold. A seminal work which studies these issues is presented by Duffie et al. \cite{DarrellD}, however, their approach relies on static beliefs and manual intervention by traders and is not automated.

More recently, agent lenders have turned to building machine learning based models to set lending rates. While more scalable and adaptive than rule-based approaches, there are still limitations. The first is that there is no ground truth label available for training, as the “correct” lending rate is unobservable, which can make training a standard supervised model difficult. The second, and perhaps most important, is that a fixed model still may not be able to adapt to changing market conditions.

The goal of this work is to explore the use of automated, dynamic, and data-driven methods for optimizing the actions of an agent lender in the SL market. Unlike the typically studied markets that use a continuous-time double auction mechanism and limit order book to handle order flows and execution in a centralized fashion, the SL market is organized more like a standard e-commerce platform where transactions occur between a buyer and seller directly. Given this important difference, existing automated trading approaches in the quantitative finance literature designed for markets that follow a limit order book structure may not be appropriate for the SL market. However, the similarities between the SL market and e-commerce platforms hint that approaches for dynamic pricing problems may be effective.

A well-studied and successful paradigm for approaching dynamic pricing problems, is the contextual multi-armed bandit framework \cite{JonasM, VShah, MisraK}. In this framework, an agent observes context from its environment and uses that context to sequentially make decisions. Importantly, the agent quickly receives feedback on the efficacy of its past decisions and can utilize that information when making future decisions. In this paper, we demonstrate that a contextual bandit framework can be effectively applied by an agent lender in the SL market to dynamically set lending rates and improve revenue generation over standard methods. In section 2, we cover details of the SL market. In section 3, we briefly review the standard contextual bandit framework. In section 4, we describe a custom reward function to be used in the contextual bandit model. Finally, in section 5, we train the model offline using real historical demand data and demonstrate that the contextual bandit approach can outperform traditional approaches used in the SL market.

\section{Background}
In this section, we provide a brief introduction to the securities lending market and the market mechanisms as that result from its trading infrastructure.

\begin{figure}[h]
  \centering
  \includegraphics[width=\linewidth]{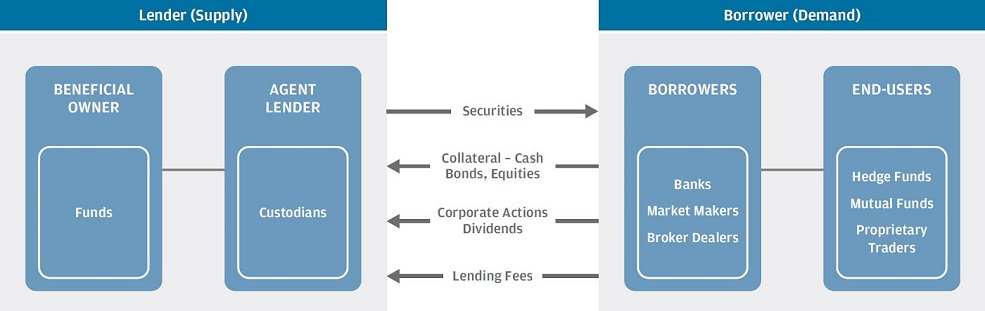}
  \caption{Interactions between entities in the securities lending process. Source: J.P. Morgan Asset Management}
  \Description{Securities lending process.}
\end{figure}

\subsection{Securities Lending and Agent Lenders}

Securities lending works by allowing a fund to temporarily lend securities that it owns to an approved borrower in return for a fee. The borrower is required to provide sufficient collateral, in the form of either cash or securities, to compensate the fund if the borrower fails to return the loaned securities in the agreed timeframe, subject to certain counterparty and liquidity risks \cite{jpmam}. Figure 1 captures the key market participants and shows how the securities lending process works.

Agent lenders play an important role in the lending market. On one hand, they lend out securities on behalf of asset owners/managers to create incremental alpha. On the other hand, they offer broker dealers a way to mitigate the consequence of frictions inherent to OTC markets at the cost of charging a lending fee \cite{DarrellD}. To reduce such cost a borrower often reaches out to multiple lenders to communicate interests for securities and borrow from the lender that offers the lowest fee, especially for hard-to-borrow names. The lending fee dependents heavily on market supply-demand dynamics. From an agent lender’s perspective, suboptimal pricing decisions lead to reduction in its own revenue and in some cases sends the wrong pricing signals to other agency lenders and distort price in the market. 

\subsection{Trading Infrastructure and Market Mechanism}

Trading in the SL markets happens on marketplace platform like the \textbf{Next Generation Trading (NGT)}\footnote{NGT is a multi-asset trading platform available 24 hours a day for real-time securities lending trading between lenders and borrowers. As of May 2023, \$113.5bn notional on average traded on NGT each day. https://equilend.com/services/ngt/.}. On the supply side, lenders actively update executable inventory by publish <Offer rate, Inventory> on \textbf{Target Availability (TA)}. On the demand side, NGT’s \textbf{Indication of Interest (IOI)} allows borrowers to reach out to multiple lenders for market discover in a non-exclusive fashion. Borrowers can also send multiple requests to the same lender with a different bid over time. Figure 2 shows an example of borrowers submitting multiple requests to one agent lender for the same security throughout the day, all of which got rejected by the lender.  After IOI, borrowers then set up a pre-defined order of lenders that they want to transact with and allow NGT to route borrow requests sequentially. 

\begin{figure}[h]
  \centering
  \includegraphics[width=\linewidth]{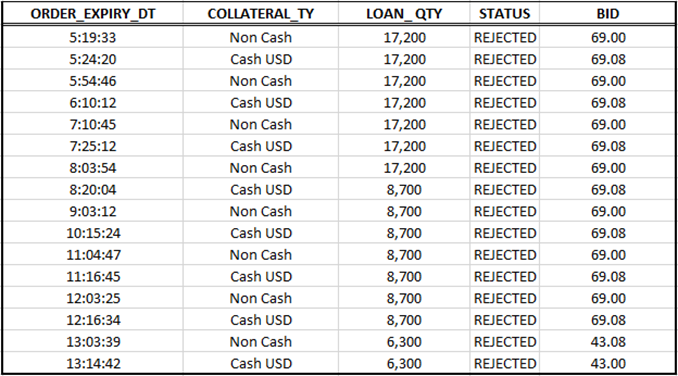}
  \caption{Borrowers sending multiple requests to a single lender for the same security with different expiration schedule}
  \Description{Request info detail from borrowers.}
\end{figure}

Trades are automatically accepted or rejected between two private parties based on few key parameters: Lender’s <Offer rate, Available inventory>, Borrower’s <Bid, Quantity requested>. A request is accepted when bid and offer agrees conditional availability, reject otherwise. Borrowers and lenders can access to transaction history on NGT at a cost. Since there’s no centralized clearing in the marketplace, there’s no notion of market clearing lending fee. The access to market transaction history allows both sides to find out where the market is trading at for price discovery\footnote{Market VWAF is the value weighted average lending fee of all outstanding loans on NGT.}. 

\subsection{Properties of a Good Pricing Strategy}

\textbf{\emph{Realistic preference.}} A necessary condition for good pricing model is that reward preference assumption should truly reflect optimization need of client/business. Poorly constructed assumptions could lead to misleading conclusions, sometimes under the disguise of complicated structure. To evaluate the realism of the assumptions, we need check if the preference is commensurable with the metrics that business cares about. For example, in SL market, the business prefers high acceptance rate of orders flow and higher price conditional on acceptance. In Section 4, we show that our preference function is designed to balance out hit rate vs ask price.  

\textbf{\emph{Responsiveness.}} Another desirable property of good pricing strategy is the responsiveness to adapt to market dynamics: the model should include parameters that explicitly captures the market dynamics allowing it to response to the change directly.  For example, when an agent lender owns a dominating share of the market, the agent lender has tremendous pricing power, the lending fee tends to be more aggressive than when the agent lender has low market share. In section 5, we provide empirical evidence showing that our reward function and market share features can capture such pricing power dynamic due to monopoly power.

\section{Contextual Bandit Problem Formulation \& Related Work}
In this section, we define the K-armed contextual bandit (CB) problem formally, and as an example, show how it can model the agent lenders’ offer rate pricing problem. We then discuss related works and their limitations.

\subsection{Contextual Bandit Formulation for an Agent Lender}
In a competitive market, the revenue optimization problem for the representative agent lender in the marketplace takes market demand as given.  Hence, it is safe to assume such lender is not likely to have any systematic impact on aggregate demand. Following previous work of Li et al. \cite{LihongL}, we model the agent lender pricing problem as a bandit problem with contextual information. Formally, a CB algorithm learn through repeated interaction over \emph{T} rounds. At each round \emph{t = 1, 2, … T}:

\begin{itemize}
\item The algorithm observes the current market environment \(x_t \in X\) where \(X\) is a set of market features the vector \(x_t\) summarizes information about market supply and demand dynamics in our setting.

\item Based on previous observed payoffs and the context, the algorithm chooses an action \(a_t \in A\), from a set of actions, \(A\). In our setting each action corresponds to a price level. Given an action, \(a_t\), we receive a reward, \(r_t(x_t, a_t)\).

\item The algorithm then improves the action selection strategy with this new observation. Note that the reward is only observed for the chosen \(a_t\) at any time step \(t\).
\end{itemize}

The CB algorithm is equipped with a set of policies \(\Pi \subseteq \{X \rightarrow A\}\), which describe how actions are chosen given the observed context. The objective of a CB algorithm is to learn a policy \(\pi \in \Pi\) which minimizes,

\begin{equation}
    Regret(\pi)=\sum_{t=1}^{T} [r_t(x_t,a_t^*)-r_t(x_t,\pi(x_t))]
\end{equation}

where \begin{equation}
    \pi(x_t) = \arg\max_{a_t}r_t(x_t,a_t)
\end{equation}

The optimal policy \(\pi^*\) is thus: \begin{equation}
    \pi^*=\arg\min_{\pi}Regret(\pi)
\end{equation}

\subsection{Related Work}
Dynamic pricing strategies for financial market has been extensively studied in the literature.  Duffie et al. \cite{DarrellD} use traditional statistical methods to dynamically determine the price, fees, and the interests rate in securities lending. The lending fee dynamics is driven by the difference in agent believes and bargaining. Their styled assumptions on believes are static and rarely hold in securities lending market as more than 80\% of the transactions are low touch transactions that settles without traders’ intervention. 

More recently, reinforcement learning (RL) has emerged as a powerful paradigm for sequential decision-making problems, where an agent learns to interact with an environment to maximize a cumulative reward. Khraishi et al. \cite{RaadK} applied an offline policy evaluation onto dynamic pricing for consumer credit with the classic Q-learning paradigm. There is plenty of past work that has shown - assuming an infinite data stream - unbiased or low-biased evaluation methods, such as \cite{10.1145/1935826.1935878, dudik2012sampleefficientnonstationarypolicyevaluation, 10.1214/14-STS500}. There is also work focusing on empirically well-performing evaluation methods, yet with a lack of unbiasedness \cite{pmlr-v32-mary14}. Li et al. \cite{LihongL} used contextual bandit techniques to make personalized recommendation with the reward function of article click through rate. Most RL methods rely on predefined reward functions, such as immediate profit or transactional success rate. Khraishi et al. \cite{RaadK} took expected immediate profit as the reward for dynamically pricing.   Such reward functions can be challenging to define accurately and may not fully capture the nuances of the field.

\section{Reward Function \& Evaluation}
In this Section, we define a bounded reward function for the agent lender and show that it can be broken down into intuitively explainable components. We then go over the evaluation methodology.  

\subsection{The Agent Lender's Reward Function}
To effectively use the contextual bandit framework, a sensible reward metric is critical. Given the nature of the securities lending market the reward function that is used should meet two criteria. First, it should prefer to lend out (or book) a security over not lending out a security. Second, if a security is lent out, it should prefer to lend out the security at as high of a rate as possible. Therefore, we define the reward as the product of two terms – booking preference and booking practice – and refer to the overall reward as revenue propensity.

\textbf{\emph{Reward}}. In a nutshell the agent lender’s reward should prefer booking over rejection and if a booking has been made, it should prefer a higher lending fee. Hence, we define the reward as the product of two parts and call it \(Revenue Propensity \in [0,1]\):

\begin{equation}
    Revenue Propensity=r_t(x_t(Bid_t^s),a_t^s,\delta)
\end{equation}

\begin{equation}
    Revenue Propensity\equiv Booking Ref*Booking Status
\end{equation}

\textbf{\emph{Booking}}. A continuous value bounded between 0 and 1. It represents the preference of a match or no match for any request \(s\).  

\(Booking Preference \in [0, 1]\) is defined as:

\begin{equation}
    bp(Bid_t^s, a_t^s)=\begin{cases}
        0, & \text{if} \  Bid_t^s<a_t^s \\
        \frac{a_t^s}{Bid_t^s}, & \text{otherwise}
    \end{cases}
\end{equation}

where \(Bid_t^s\) is the bid from the borrower and \(a_t^s\) is the proposed lending fee from the lender. Intuitively, when there is no match i.e., when \(Bid_t^s\) is below \(a_t^s\), the booking preference is zero. When a match happens, the booking preference rewards an action \(a_t^s\) that is closer to \(Bid_t^S\). Booking preference is intuitively maximized when, \(a_t^s=Bid_t^s\) , as this reflects a scenario in which the full extent of the borrower’s willingness to pay has been met. Figure 3 plots the value of booking preference for various level of \(a_t^s\) as a function of bid to illustrate the tradeoff between hit rate and high booking preference value.

\begin{figure}[h]
  \centering
  \includegraphics[width=\linewidth]{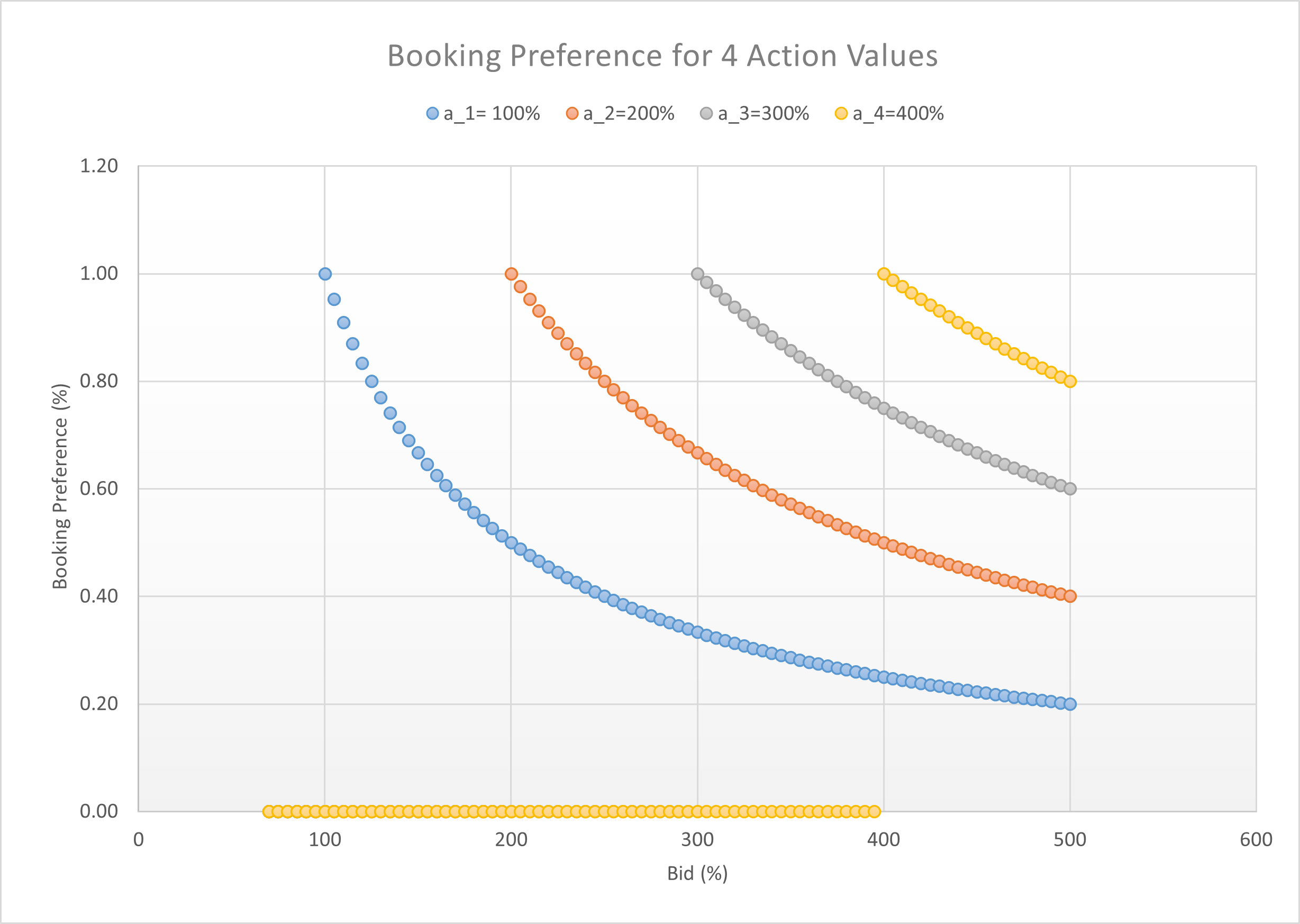}
  \caption{Booking preference as a function of bid for various \(a_t^s\).}
  \Description{Booking preference as a function of bid for various \(a_t^s\).}
\end{figure}

\textbf{\emph{Booking Status}}. A binary variable reflecting a match or not. \(Booking Status \in \{0,1\}\) is defined as follows:

\begin{equation}
    I_b(Bid_t^s,a_t^s,\delta)=\begin{cases}
        0, & bp(Bid_t^s, a_t^s)=0 \ \text{or} \ Bid_t^s < \delta C \\
        1, & \text{otherwise}
    \end{cases}
\end{equation}

where \(\delta\) is a threshold multiplier and \(C\) is a benchmark value to detect spoofing requests, which occurs when a borrower tries to book a trade with an extremely low Bid to influence the market. Booking Status can be estimated via any model capable of producing binary predictions. 
Thus, the overall reward for an action from a policy \(\pi\) given a context at time \(t\) is \(r_t(x_t,\pi(x_t))\) and the reward for an oracle-provided (i.e., the optimal) action \(a_t^*\) is \(r_t(x_t,a_t^*)\). Finally, our optimization problem is to minimize the total regret, which is defined as the difference in the oracle-provided expected revenue and the expected revenue generated by a policy. The expected revenue of a trade can be derived by multiplying the reward (i.e., \(Revenue Propensity\)), the market value of the loan (\(MarketVal\)), and the lending fee (a percentage of the market value of the intended security, in this case, the price of the oracle-provided action \(a_t^*\)).

\begin{equation}
    Regret(\pi)=\sum_{t=1}^{T}[r_t(x_t,a_t^*)-r_t(x_t,\pi(x_t))]*MarketVal*a_t^*
\end{equation}

\begin{equation}
    \pi^*=\arg\min_{\pi}Regret(\pi)
\end{equation}

\subsection{Offline Evaluation}
Evaluation for problems with sequential interactions are always tricky. In our case, we want to measure the cumulative reward for a given policy. The ideal situation is to have each policy test in an ‘online’ on the trading platform, which is not feasible in our research setting. Hence, we followed the unbiased offline evaluation method in Li et al. \cite{LihongL} by using historical logged transaction details. We built a simulator to model the bandit process from the logged transaction data, and then evaluated each policy using the following steps:

\begin{enumerate}
    \item Initialize the action vector and/or matrix for each action arm
    \item Observe a booking request and create the context vector using the market supply-demand signals 
    \item Estimate the Booking Status or Revenue using different strategies (elaboration on the strategies could be found below) by taking the context vector and action vectors/matrices
    \item Choose the action arm that gives the highest reward (i.e., estimated revenue)
    \item Update the parameters of the vectors/matrices of the action arm that gives the highest estimated revenue
    \item Keep looping through step 2, 3, 4, and 5 to balance exploration and exploitation dynamically
\end{enumerate}

In our experiment we tested 4 exploration/exploit strategies:

\begin{itemize}
    \item \textbf{Linear Upper Confidence Bound (LinUCB. \emph{With a direct estimation on the Revenue.}):} We use the originally LinUCB presented in Li et al \cite{LihongL}, This method creates a confidence interval around the estimated mean reward of an action and chooses the action with the highest upper confidence bound. In this strategy, the reward is just the revenue of a booking request since there is no need to estimate the booking preference.
    \item \textbf{Regularized Logistic Regression [Algorithm 1]:} This strategy initializes two parameters $m$ - which is either the parameters (if not sampling) or the mean of the parameters (if sampling) - and $q$ - which is the variance (if sampling, and will be ignored if not sampling) - for each action arm \(a\) to estimate a reward \(\hat{r_t}(x_t,a)\), which is the \textbf{\emph{Booking Status}} in this case, by applying a sigmoid operation on the context \(x_t\) and $m$ (if not sampling) or $\mathcal{N}(m, q)$ (if sampling), with the following variations:
    \begin{itemize}
        \item Upper Confidence Bound (LRUCB): \textbf{\emph{Booking Status}} is estimated by inputting the dot product of the context $x_t$ and the parameters $m$ into a sigmoid function with a predefined confidence bound factor $\alpha$. \textbf{\emph{This is not a sampling method.}}
        \item Thompson Sampling (LRTS): \textbf{\emph{Booking Status}} is estimated by inputting the dot product of the context $x_t$ and a normal sampling (mean $m$, variance $q$) into a sigmoid function. \textbf{\emph{This is a sampling method.}}
        \item Epsilon Greedy (EG): \textbf{\emph{Booking Status}} is estimated by inputting the dot product of the context $x_t$ and the parameters $m$ into a sigmoid function. \textbf{\emph{This is not a sampling method.}}
    \end{itemize}
\end{itemize}

LRUCB, LRTS, and EG strategies use the same algorithm backbone \textbf{[Algorithm 1]} of logistic regression to estimate the value of Booking Status (for EG) or the mean value of Booking Status (for LRUCB and LRTS).    Once the booking status is estimated, we can estimate the reward of an action by multiplying it with the booking preference and the market value of a certain security.

\begin{algorithm}
\caption{Regularized Logistic Regression Algorithm for Reward Estimation}\label{alg:cap}
\begin{algorithmic}
    \State $m \gets \vec{0}$ \Comment{\emph{Initiate the mean of the parameters (if sampling) or the parameters (if not sampling)}}
    \State $q \gets \lambda\vec{1}$ \Comment{\emph{Initiate the variance of the parameters (if sampling)}}
    \State $\alpha \gets \text{Confidence bound factor for each one of the action arms.}$
    \For{$t \in T$}
    \For{$\forall\alpha \in A_t$}
    \State $x_t \gets \text{Observe a new context}$
    \If{using LRUCB}
        \State $\mu \gets Sigmoid(m, x_t)$ \Comment{\emph{/*Estimate mean of the reward*/}}
        \State $ \hat{r_t}(x_t,a) \gets UCB(\mu, a)$ \Comment{\emph{/*Add upper bound to the mean of the reward*/}}
    \ElsIf{using LRTS}
        \State $p \gets Thompson \ Sampling$ \Comment{\emph{/*Normal sample to estimate the parameters*/}}
        \State $ \hat{r_t}(x_t,a) \gets Sigmoid(p, x_t)$ \Comment{\emph{/*Then estimate the reward*/}}
    \ElsIf{using EG}
        \State $ \hat{r_t}(x_t,a) \gets Sigmoid(m, x_t)$ \Comment{\emph{/*Directly estimate the reward*/}} 
    \EndIf
    \EndFor
    \If{$a_t=\text{current booking request's assigned action}$}
        \State $\text{update} \ m\ $
        \State $\text{update}\ q\ \text{(if sampling)}$
    \EndIf
    \EndFor
\end{algorithmic}
\end{algorithm}

\section{Experiment}
In this section, we use the offline evaluation method of Li et al. \cite{LihongL} to compare the 4 contextual bandit algorithms with 4 non-contextual bandit pricing strategies. We describe the experiment setup briefly followed by discussion of results.

\subsection{Experiment Setup}
This subsection gives a detailed description of our experimental setup, including data, contextual feature construction, performance evaluation, and competing algorithms.

\subsubsection{\textbf{Data}}
We trained our models with historical bid-level data from NGT autoborrow, so that each observation contains a timestamp, quantity, bid, and accept or reject status. Additional data for market features including market share, utilization, other source of supply, borrower’s bid from the demand side is obtained separately from other sources. For the scope the experiment, we focus on the U.S. market and securities with a lending fee in the range from 1\%- 10\% as the securities with very high lending fee will be manually trader by the desk and the GC securities will be priced at 25 bp in general. 

\subsubsection{\textbf{Contextual Feature Construction}}
Our context vector consists of features describing market signals, the agent lender’s information, and demand side information. We use 4 market supply-demand signals and 1 demand feature including (1) \textbf{\emph{Utilization $\in [0, 1]$}}, (2) \textbf{\emph{Agent Lender Market Share $\in [0, 1]$}}, (3) \textbf{\emph{Alternative Supply Signal $\in [0, 1]$}}, (4) \textbf{\emph{Return Signal $\in [0, 1]$}}, (5) \textbf{\emph{BID Signal Scaled $\in [0, 1]$}}.

\begin{enumerate}
    \item \textbf{\emph{Utilization $\in [0, 1]$}} stands for the market level demand to supply ratio
    \item \textbf{\emph{Agent Lender Market Share $\in [0, 1]$}} is a proxy for the market power of the agent lender
    \item \textbf{\emph{Alternative Supply Signal $\in [0, 1]$}} is a proxy for the alternative source of supply outside NGT
    \item \textbf{\emph{Return Signal $\in [0, 1]$}}, which is also called \textbf{\emph{Return Over Notional}} is a proprietary demand signal for the agent lender 
    \item \textbf{\emph{BID Signal Scaled $\in [0, 1]$}} is the ratio of BID to EWMA, which is a bid level feature used to gauge borrower interest level
\end{enumerate}

A good contextual feature should discriminate various market condition, i.e., we should expect to see reward ranking amongst actions to vary over the range of our context. For example, in Figure 4, we plot the average Booking Preference of 4 type of pricing strategies in our action space vs 2 contextual features -market share and other sources of supply.  The top chart shows that the action with highest booking preference varies across different market share levels, i.e., not a single action pareto dominates the rest. The best action in mid-range (40-60\%) might not be the best action in high market power situation (>70\%). Moreover, it shows that our booking preference penalizes pricing strategy \(P_{rule}^s(t)\) that is overly aggressive, but less so when market share is > 85\%, in which the agent lender will have monopoly pricing power.

\begin{figure}[h]
  \centering
  \includegraphics[width=\linewidth]{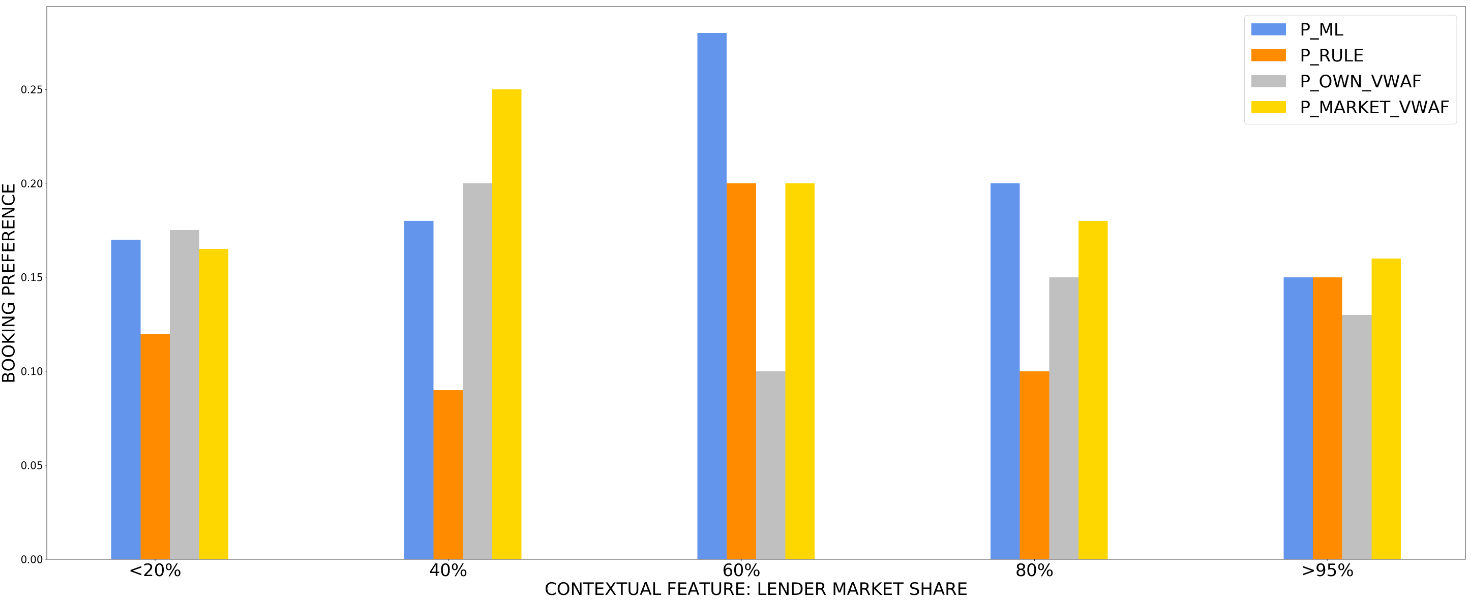}
  \includegraphics[width=\linewidth]{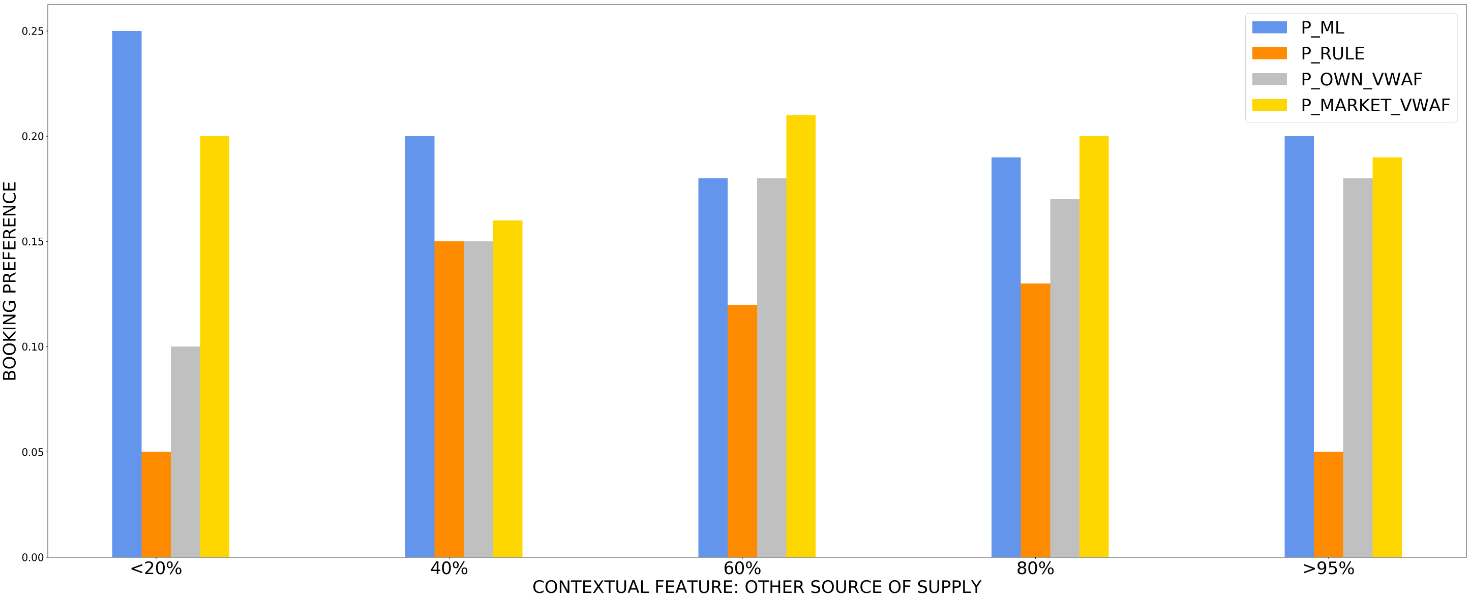}
  \caption{Comparison of average Booking Preference of 4 actions given the context (x axis - market share (top), other source of supply (bottom), y axis - Booking Preference).}
  \Description{Booking preference given different context.}
\end{figure}

\subsection{Method Comparisons}
We compared contextual bandits-based policies with other non-CB policies commonly used by agent lenders. We estimated revenue generated by 8 policies trained below. 

\textbf{I. Non-CB Pricing Policies.} These are the pricing strategies that commonly used by major agent lenders. The strategies span our action space for CB setup:

\begin{itemize}
    \item \textbf{$P_{ML}^s(t)$: } ML based supervised learning predictive pricing policy 
    \item \textbf{$P_{rule}^s(t)$: } Agent Lender rule-based pricing policy 
    \item \textbf{$P_{own\ vwaf}^s(t)$: } Agent Lender’s existing book VWAF 
    \item \textbf{$P_{market\ vwaf}^s(t)$: } Market VWAF of all agent lenders 
\end{itemize}

\begin{figure}[h]
  \centering
  \includegraphics[width=\linewidth]{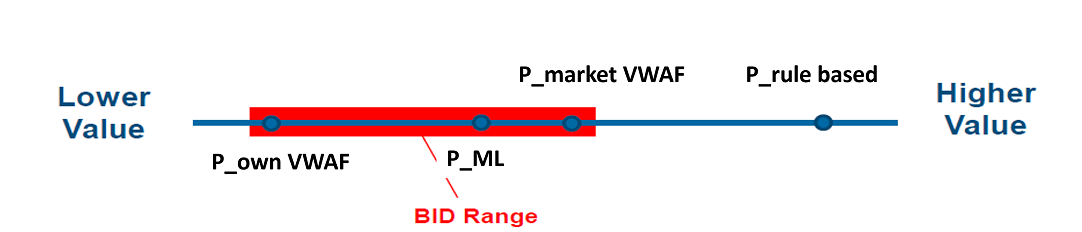}
  \caption{Cardinal ranking of aggressiveness of 4 type of pricing strategies that in the action space.}
  \Description{Cardinal ranking of aggressiveness of 4 type of pricing strategies that in the action space.}
\end{figure}

Figure 5 shows the cardinal ranking of aggressiveness of the 4 actions in our experiment, from low to high value are \textbf{$P_{own\ vwaf}^s(t)$}, \textbf{$P_{ML}^s(t)$}, \textbf{$P_{market\ vwaf}^s(t)$}, and lastly \textbf{$P_{rule}^s(t)$}. The higher the rate, the more aggressive the agent lender is in its pricing strategy.

\textbf{II. CB Pricing Policies}. As mentioned in subsection 4.2, the CB pricing strategies are:
\begin{itemize}
    \item LinUCB
    \item LRUCB
    \item Epsilon-Greedy (EG)
    \item Thompson Sampling (LRTS)
\end{itemize}

\subsection{Assumption and Setup Explanations}

Financial market is known for its nature of tumultuous signals, and in this work, we had to make certain assumptions and confine the scope the problem, in order to ensure the overall implementation. We believe that there are 2 items that are worth mentioning regarding our experiments.

\textbf{I. Non-stationarity.} In this work, we use historical market supply-demand data as context to evaluate multiple contextual bandit strategies. Yet, in reality with real-time requests, it’s nearly impossible to obtain such data. In addition, market conditions often change rapidly and unexpectedly, which potentially forms a non-stationary environment. Hence, the offline nature of this experiment might face quite an uncertainty in real-time applications. However, the setup and implementation of this experiment are designed for a fair evaluation on the different yet related approaches – with a focus on how different models react and adapt to the securities lending market – serving as a corner stone for exploring further possibilities of contextual bandit, or even the broader concept of reinforcement learning, in this market.

\textbf{II. Differences among agent lenders.} We mainly focus on how one agent lender could best setup their pricing strategy internally, where they only consider the available market data along with the prices. Despite the fact that we are aware of the potential outcome of a price war if not taking the specialties of the agent lenders into account, in this work, we direct our attention more onto the question of whether the proposed contextual bandit-based approaches could perform better than other existing approaches.

\subsection{Results and Discussion}
To test each pricing strategy, we use 5-day worth of trading log data from NGT. The first 4 days are used for training, and the last day is used for testing. We measure the total estimated revenue in a 5-day sliding window fashion (i.e., approximately the first 4 days will be used for training and the last day will be used for testing), to test the consistency of the results.

\begin{table*}[t]
\begin{tabular}{|cccccccccc|}
\hline
\multicolumn{10}{|c|}{\textbf{Estimated Revenue of the Last Day with a 5-Day Period in selected date in 2021 and 2023 (Millions USD)}}                                                                                                                                                                                                                    \\ \hline
\multicolumn{1}{|c|}{\multirow{2}{*}{Policy}} & \multicolumn{3}{c|}{2021}                                                 & \multicolumn{3}{c|}{2021}                                                 & \multicolumn{3}{c|}{2023}                            \\ \cline{2-10} 

\multicolumn{1}{|c|}{}                  & \multicolumn{1}{c|}{Apr 21-27} & \multicolumn{1}{c|}{Apr 22-28} & \multicolumn{1}{c|}{Apr 23-29} & \multicolumn{1}{c|}{Nov 16-22} & \multicolumn{1}{c|}{Nov 17-23} & \multicolumn{1}{c|}{Nov 18-24} & \multicolumn{1}{c|}{Apr 27-May 3} & \multicolumn{1}{c|}{Apr 28-May 4} & May 1-5 \\ \hline

\multicolumn{1}{|c|}{**ML-Based}         & \multicolumn{1}{c|}{1.14} & \multicolumn{1}{c|}{1.5} & \multicolumn{1}{c|}{1.98} & \multicolumn{1}{c|}{1.65} & \multicolumn{1}{c|}{0.82} & \multicolumn{1}{c|}{3.1} & \multicolumn{1}{c|}{0.39} & \multicolumn{1}{c|}{0.43} &  1.65 \\ \hline

\multicolumn{1}{|c|}{**Rule-Based}       & \multicolumn{1}{c|}{0.41} & \multicolumn{1}{c|}{0.53} & \multicolumn{1}{c|}{1.05} & \multicolumn{1}{c|}{0.45} & \multicolumn{1}{c|}{0.35} & \multicolumn{1}{c|}{2.39} & \multicolumn{1}{c|}{0.10} & \multicolumn{1}{c|}{0.34} & 0.28 \\ \hline

\multicolumn{1}{|c|}{*LinUCB}           & \multicolumn{1}{c|}{0.9} & \multicolumn{1}{c|}{1.2} & \multicolumn{1}{c|}{1.88} & \multicolumn{1}{c|}{1.41} & \multicolumn{1}{c|}{0.78} & \multicolumn{1}{c|}{3.3} & \multicolumn{1}{c|}{0.86} & \multicolumn{1}{c|}{0.56} & 2.2 \\ \hline

\multicolumn{1}{|c|}{*LRUCB}          & \multicolumn{1}{c|}{1.64} & \multicolumn{1}{c|}{1.61} & \multicolumn{1}{c|}{2.56} & \multicolumn{1}{c|}{1.81} & \multicolumn{1}{c|}{1.32} & \multicolumn{1}{c|}{4.25} & \multicolumn{1}{c|}{1.07} & \multicolumn{1}{c|}{0.75} & 2.43 \\ \hline

\multicolumn{1}{|c|}{*EG}     & \multicolumn{1}{c|}{1.61} & \multicolumn{1}{c|}{1.56} & \multicolumn{1}{c|}{2.5} & \multicolumn{1}{c|}{1.76} & \multicolumn{1}{c|}{1.23} & \multicolumn{1}{c|}{4.15} & \multicolumn{1}{c|}{1.06} & \multicolumn{1}{c|}{0.72} & 2.21 \\ \hline

\multicolumn{1}{|c|}{*LRTS}   & \multicolumn{1}{c|}{1.65} & \multicolumn{1}{c|}{1.62} & \multicolumn{1}{c|}{2.57} & \multicolumn{1}{c|}{1.86} & \multicolumn{1}{c|}{1.33} & \multicolumn{1}{c|}{4.26} & \multicolumn{1}{c|}{1.07} & \multicolumn{1}{c|}{0.76} & 2.44 \\ \hline

\multicolumn{1}{|c|}{**Market VWAF}         & \multicolumn{1}{c|}{1.38} & \multicolumn{1}{c|}{1.39} & \multicolumn{1}{c|}{2.07} & \multicolumn{1}{c|}{1.61} & \multicolumn{1}{c|}{0.81} & \multicolumn{1}{c|}{3.86} & \multicolumn{1}{c|}{0.92} & \multicolumn{1}{c|}{0.48} & 2.35 \\ \hline

\multicolumn{1}{|c|}{**Lender VWAF}          & \multicolumn{1}{c|}{0.85} & \multicolumn{1}{c|}{1.24} & \multicolumn{1}{c|}{1.6} & \multicolumn{1}{c|}{1.13} & \multicolumn{1}{c|}{1.08} & \multicolumn{1}{c|}{3.16} & \multicolumn{1}{c|}{0.48} & \multicolumn{1}{c|}{0.49} & 0.96 \\ \hline

\end{tabular}
\caption{Contextual bandit-based lending fee optimization experiment: comparing the simulated revenue based on sample of 5-day NGT demand data for 8 distinct pricing strategies. Contextual bandit-based policies are denoted with a (*), while the traditional, non-contextual bandit-based approaches are denoted with a (**). (The anti-spoofing factor $\delta C$ is set to 0.85 in all experiments).}
\end{table*}

Table 1 shows the revenue generated by each policy based on our simulation using real demand data from 3 consecutive 5-day periods April 2021 and another 3 from November 2021. The following patterns can be seen quite consistently in our results:

The first, and most encouraging, is that the contextual bandit-based methods almost uniformly outperform the traditional non-contextual bandit pricing policies. This is perhaps unsurprising, as the contextual bandit approach can use past and current market conditions to determine which pricing strategy is likely to be most effective given the current environment.

The second is that the contextual bandit approaches provide insight into which traditional approaches are most effective, and why. Table 2 shows a ratio of how often each pricing strategy was chosen by each contextual bandit method. From this, we can see that the agent lender’s Own VWAF rate and ML-based rate were selected most often. This likely occurs because the agent lender’s Own VWAF rates are usually lower than the Bid and the other rates are usually higher than the Bid. Hence, on average, there would be more match between the Bid and the agent lender’s Own VWAF rates. On the other hand, when the Bid are greater than ML-based rates, the ML-based rates would be very close to Bid, hence, under this scenario, picking ML-based rates gives higher booking preference for individual incoming booking requests, rendering higher total revenue.

\begin{table}[]
\begin{tabular}{|c|ccc|}
\hline
\multirow{2}{*}{Policy} & \multicolumn{3}{c|}{2023 May 1-5}                                                    \\ \cline{2-4}

                  & \multicolumn{1}{c|}{Own VWAF} & \multicolumn{1}{c|}{Market VWAF} & \multicolumn{1}{c|}{ML/Rule-Based} \\ \hline
                  
       *LinUCB           & \multicolumn{1}{c|}{0.84} & \multicolumn{1}{c|}{0.03} & 0.13 \\ \hline
       
       *LRUCB           & \multicolumn{1}{c|}{0.32} & \multicolumn{1}{c|}{0.23} & 0.45  \\ \hline
       
       *EG           & \multicolumn{1}{c|}{0.24} & \multicolumn{1}{c|}{0.24} & 0.51  \\ \hline
       
       *LRTS           & \multicolumn{1}{c|}{0.26} & \multicolumn{1}{c|}{0.24} & 0.50  \\ \hline
\end{tabular}
\caption{Ratio of each action selected as the best action in 2023 May experiments. (Note: Rule-Based ratios are negligibly small, hence are merged into ML-Based ratios.)}
\end{table}

From Table 2, we see that the contextual bandit approaches powered by a regularized logistic regression backbone for the \textbf{\emph{Revenue Propensity}} (i.e., LRUCB, EG, and LRTS) are able to capture the niche of the market dynamics. With LinUCB, the model tend to pick agent lender’s Own VWAF rates, which intuitively makes sense because they are usually the lowest. However, LRUCB, EG, and LRTS seem to be able to better leverage the market signals, as they would pick ML-based rates when the Bid is just slightly higher than ML-based rates – along with favorable market conditions (i.e., when a good \textbf{\emph{Revenue Propensity}} occurs) – and this is likely the rationale behind the higher selection ratio in ML-based rates. Additionally, in empirical evaluation, LRUCB, EG, and LRTS indeed generate higher estimated revenue, with LRTS leading the list.

Finally, among the contextual bandit frameworks we investigated, the policies which utilized our custom reward function by estimating the booking status using logistic regression performed better than the framework which estimated revenue directly. This likely occurs because without incorporating the booking status, it is implicitly assumed that a price will be accepted, which is likely too aggressive of an assumption and thus leads to a disproportionate amount of missed bookings.

\section{Conclusion}
In this paper we investigated the use of the contextual bandit framework in improving the ability of an agent lending to optimize their pricing in the securities lending market. Using real historical demand data from the securities lending market, we benchmarked the performance of optimal prices chosen by common contextual bandit frameworks. We demonstrated that contextual bandit frameworks can be effectively used to choose between static prices recommended by supervised learning model as well as three logic-based pricing rules which are commonly used in the industry. In our historical data-based simulation, we have shown that our contextual bandit framework can learn on historical data to recommend an optimal price, without the need of some heavy model assumption and computation as in any supervised learning model. 

We also demonstrated that our models improve previous state-of-art solutions, providing higher PnL than any single pricing strategies. We discussed the main advantages of the Contextual Bandit framework, and we demonstrated that it is flexible enough to consider multiple pricing strategies at once. For future work, we would like to explore the possibility of embedding the input context into latent space for capturing more sophisticated features, improve the model performance on a few directions, such as evaluation with real time demand data (i.e., online evaluation), which requires a more sophisticated consideration on infrastructure readiness in a production setting, and tackling the items mentioned in section 5.3, which will make our work more comprehensive.

\section{Disclaimer}
This paper was prepared for informational purposes in part by the Quantitative Research group of JPMorgan Chase \& Co. and its affiliates (“J.P. Morgan”) and is not a product of the Research Department of J.P. Morgan. J.P. Morgan makes no representation and warranty whatsoever and disclaims all liability, for the completeness, accuracy or reliability of the information contained herein. This document is not intended as investment research or investment advice, or a recommendation, offer or solicitation for the purchase or sale of any security, financial instrument, financial product or service, or to be used in any way for evaluating the merits of participating in any transaction, and shall not constitute a solicitation under any jurisdiction or to any person, if such solicitation under such jurisdiction or to such person would be unlawful.

\bibliographystyle{ACM-Reference-Format}
\bibliography{ref}










\end{document}